\documentclass{edp-conf}
\usepackage{graphicx}
\renewcommand{\d}{\mathrm{d}}

\begin{document}

\TitreGlobal{SF2A 2002}
\title{Cosmic Magnification} 
\author{Brice M\'enard~$^{1,}$}
\address{Max-Planck-Institut f\"ur Astrophysik, P.O.~Box 1317,
  D--85741 Garching, Germany}
\address{Institut d'Astrophysique de Paris, 98 bis Bld Arago, F--75014,
  Paris, France}
\runningtitle{Cosmic Magnification}
\setcounter{page}{237}
\index{Brice M\'enard}

\maketitle
\begin{abstract} 

I present the  current status of the \emph{cosmic magnification} produced by 
systematic amplification of background sources by large-scale structures.
After introducing its principle, I focus on  its interests
for cosmology and underline its
complementary aspect to cosmic shear and galaxy auto-correlations.
I finally discuss recent investigations using higher-order statistics.

\end{abstract}

\section{The principle}

The large-scale distribution of matter in the Universe induces
systematic lensing effects on background objects like quasars or
distant galaxies. These lensing effects are of two kinds: the apparent
shape of background sources is distorted and their flux is amplified
or desamplified.  Except for strong configurations that produce image
multiplications and local lensing events, in general the lensing
signal is weak, so analyses must be carried out over wide
fields-of-view in order to measure statistically weak distortion of
shapes and/or weak flux variations of lensed objects.  The statistical
properties of cosmic shear and cosmic magnification signals can then
be used to probe cosmological parameters and the power spectrum of
density fluctuations in the Universe.\\ 
Cosmic shear can be measured
by analysing the correlations between the shapes of background
galaxies.  Likewise, for cosmic magnification the cumulative
amplifications of background sources by foreground matter gives rise
to angular correlations between two populations that are physically
uncorrelated.  This can be observed by considering two populations at
different distances, like foreground and background galaxies or
foreground galaxies and background quasars.  Focusing on the latter
for example, the expression of the corresponding lensing-induced
correlation reads (Bartelmann, 1995):
\begin{eqnarray}
  w_\mathrm{QG}(\theta)&=&\frac
  {\left\langle[n_\mathrm{Q}(\vec\phi)-\bar{n}_\mathrm{Q}]
    [n_\mathrm{G}(\vec\theta+\vec\phi)-\bar{n}_\mathrm{G}]
    \right\rangle}{\bar{n}_\mathrm{Q}\,\bar{n}_\mathrm{G}}\;,\nonumber\\
  &=&2\,(\alpha-1)\,\bar b(\theta)\,w_{\kappa\delta}(\theta)\;.
\end{eqnarray}
Where $n_\mathrm{Q}$ and $n_\mathrm{G}$ are respectively the number densities 
of quasars and galaxies,  and $\alpha$ is the observed
slope of the quasar number counts as a function of magnitude.
$\bar b$ denotes the galaxy bias, averaged over all galaxies of the
sample.
The cross-correlation $w_{\kappa\delta}(\theta)$ is related to a projection of
the dark matter power spectrum:
\begin{equation}
  w_{\kappa\delta}(\theta)=
  \int\d w\,
  \frac{p_\kappa(w)\,p_\delta(w)}{f_K^2(w)}\,\nonumber\\
  \int\frac{s\d s}{2\pi}\,
  P_\delta\left(\frac{s}{f_K(w)},w\right)\,\mathrm{J}_0(s\theta)\;.  
\end{equation}
where $w$ is the radial comoving distance along the line-of-sight,
$p_\delta(w)$ describes the galaxy distribution
and $p_\kappa(w)$ is the lensing efficiency of the quasar distribution
(see M\'enard \& Bartelmann 2002 for more detail). This cross-correlation
is plotted in Fig. 1 for a standard cosmological model. It shows that
lensing creates an overdensity of quasars around foreground galaxies
at the 2\%-level.\\
During the past decade, many people attempted to detect this
cosmological signal, mainly by measuring quasar-galaxy correlations
from different surveys (see Bartelmann \& Schneider for a review).
 Their results ranged from null detections to unexpected 
  strong signals, with amplitudes  more than one order of magnitude higher 
than theoretical predictions, making this field quite confusing  
 for a long time.
The first clear detection of lensing by large-scale structures was
actually revealed by cosmic shear (van Waerbeke et al. 2000).  This
field is progressing incredibly fast, mainly due to the improvement of
panoramic CCD cameras, and is now one of the most reliable tools to
constrain $\Omega_0$, $\sigma_8$ and the dark matter power spectrum
(see Mellier et al. 2002 for a review).
In parallel, the Sloan Digital Sky Survey (SDSS) has now enough area
to permit the measurements of both quasar-galaxy and galaxy-galaxy
correlations.  Ongoing works with SDSS has for the first time
detected a signal based on quasar-galaxy correlations with an
amplitude roughly in agreement with the expectations
\footnote{This indicates that previous survey might have been not
homogeneous enough to conduct such measurements. Moreover redshift
information for all galaxies allow to cancel
contaminations by intrinsic quasar-galaxy correlations due to galaxies
located at similar redshifts than the quasars.}.  It is therefore
expected that cosmic magnification provides some constraints on both
the galaxy bias and cosmology in a near future, as cosmic shear does.

\section{Comparison to galaxy correlations and cosmic shear}

\begin{figure}[h]
   \centering
   \includegraphics[width=7cm]{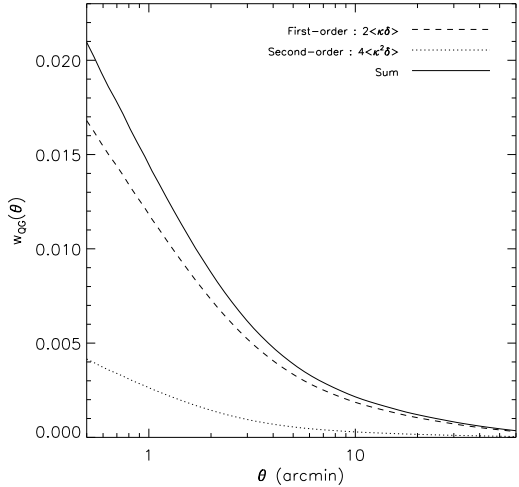}
   \caption{Expected overdensity of quasars around foreground galaxies computed
     for a standard $\Omega=0.3$, $\Lambda=0.7$ CDM cosmology with $\sigma_8=0.9$. 
     We consider here the typical values $\bar b=1$, $\alpha=2$ (valid for bright quasars) and
     use a source redshift of $z_s=2$.}
   \label{figure_mafig}
\end{figure}

The correlation functions of galaxies 
 probe both cosmology and the bias parameter of galaxies.
However, while galaxy autocorrelations measure quantities related 
to the second moment of the galaxy overdensity $\delta_\mathrm{G}$, 
cosmic magnification is a cross-correlation between
$\mu$ (or $\kappa$ in the first-order approximation: $\mu=1+2\kappa$) 
and $\delta_\mathrm{G}$. Cosmic magnification
is therefore linearly sensitive to the galaxy bias and, for a given cosmological
model, the comparison between 
   the measurements of $w_\mathrm{QG}(\theta)$ and
$w_\mathrm{GG}(\theta)$ can provide useful
  informations  the non-linearity and/or stochasticity
of the biasing. Note that this would not be possible by having one of 
these two measurements only.\\

Cosmic shear and cosmic magnification are both probing the same
physical phenomenon, namely the distorsion of light beams produced by the
cumulative effects of mass inhomogeneities in the Universe, but they
are measuring completely different quantities on the sky. Therefore,
each of these two tools has its own practical properties. 
For cosmic magnification, the main drawbacks are as follows:\\
$\bullet~~$Lensing effects due to large-scale structures are ''everywhere''
but cosmic magnification can only probe them in the vincinity of
quasars. Therefore, a huge homogeneous survey is needed. (The source density
is higher in the case of galaxy-galaxy correlations, but the slope 
$\alpha$ of the galaxy number count is closer to unity.)\\
$\bullet~~$In order to extract some useful information from the 
correlations, one has to know or assume the relation between the
galaxy and dark matter distributions, which is difficult at small
scales.\\
On the other hand, it presents some advantages compared to cosmic shear:\\
$\bullet~~$Measuring cosmic magnification can be achieved by a straightforward 
measurement,
namely counting objects (no shape measurement, no PSF correction, etc.).\\
$\bullet~~$Cosmic magnification does not use the crucial assumption that
galaxy are intrinsically randomly oriented.

\section{Higher-order statistics}

Recently, the use of higher-order statistics allowed two improvements
on the theoretical side of cosmic magnification:
M\'enard et al. (2002b) investigated the Taylor expansion of the
magnification up to second order. Including this term in the
cross-correlation between quasars and galaxies gives an additional
contribution: 
$w_\mathrm{QG}(\theta)= 2\bar b\;(\alpha-1)
\left[ \langle \kappa\kappa \rangle + \alpha\langle \kappa\kappa^2 \rangle
\right]$. 
They found this term to be non-neglegible. It can indeed contribute to
roughly 25\% of the total amplitude (Fig. 1) and becomes even more
important when the redshift of the sources and/or the slope $\alpha$
increase. Magnification related statistics always underestimated its
contribution previously.\\ In addition to the two-point quasar-galaxy
correlations, some useful informations can be extracted from measuring
a triple quasar-galaxy-galaxy correlator. As shown in M\'enard et
al. (2002a), using together these two statistics allows to directly
probe the matter density $\Omega_0$ in the angular range where the
galaxy bias has a linear behaviour.

\section{Outlook}

As surveys mapping the large-scale structure of the Universe become
wider and deeper, measuring cosmological parameters as well as the
galaxy bias with cosmic magnification will become increasingly
efficient and reliable. 
As cosmic shear has given valuable constraints on cosmological
parameters and the power spectrum, cosmic magnification will be able
to confirm these results and investigate the relation between mass and
light.
Given their specific properties, results from cosmic shear/magnification 
as well as galaxy autocorrelations can be used jointly to give interesting
results regarding the distribution of visible and dark matter.\\

\acknowledgements{This work was supported by the TMR Network
``Gravitational Lensing: New Constraints on Cosmology and the
Distribution of Dark Matter'' of the EC under contract
No. ERBFMRX-CT97-0172.}

\end{document}